\documentclass[pre,twocolumn,aps]{revtex4}
\usepackage{graphicx}
\usepackage{comment}
\usepackage{bm}
\usepackage{amsmath,amssymb} 
\usepackage{epstopdf}
\usepackage{verbatim}
\usepackage{float}

\begin{document}

\title{Thermodynamics of condensed matter with strong pressure-energy correlations}
\author{Trond S. Ingebrigtsen, Lasse B{\o}hling, Thomas B. Schr{\o}der, and Jeppe C. Dyre}
\email{dyre@ruc.dk}
\affiliation{DNRF Centre ``Glass and Time'', IMFUFA, Department of Sciences, Roskilde University, Postbox 260, DK-4000 Roskilde, Denmark}
\date{\today}

\begin{abstract}
We show that for any liquid or solid with strong correlation between its $NVT$ virial and potential-energy equilibrium fluctuations, the temperature is a product of a function of excess entropy per particle and a function of density, $T=f(s)h(\rho)$. This implies that 1) the system's isomorphs (curves in the phase diagram of invariant structure and dynamics) are described by $h(\rho)/T={\rm Const.}$, 2) the density-scaling exponent is a function of density only, 3) a Gr{\"u}neisen-type  equation of state applies for the configurational degrees of freedom. For strongly correlating atomic systems one has $h(\rho)=\sum_nC_n\rho^{n/3}$ in which the only non-zero terms are those appearing in the pair potential expanded as $v(r)=\sum_n v_n r^{-n}$. Molecular dynamics simulations of Lennard-Jones type systems confirm the theory.
\end{abstract}

\maketitle

The class of strongly correlating liquids was introduced in Refs. \onlinecite{scl_2} and \onlinecite{scl}. These liquids are defined by having a correlation coefficient above 0.9 of the constant-volume equilibrium fluctuations of virial $W$ and potential energy $U$. The $WU$ correlation coefficient varies with state point, but we found from computer simulations that a system has either poor $WU$ correlations in the entire phase diagram or is strongly correlating at most of its condensed-phase state points \cite{scl,scl_2,IV,V,rome}. Van der Waals and metallic liquids are generally strongly correlating, whereas hydrogen-bonded, ionic, and covalently bonded liquids are generally not. The solid phase is at least as correlating as the liquid phase. Theoretical arguments, numerical evidence, and experiments show that strongly correlating liquids are simpler than liquids in general \cite{scl,scl_2,IV,V,rome,ditte,simple}.

The simplicity of strongly correlating liquids compared to liquids in general \cite{liq} derives from the fact that the former have ``isomorphs'' in their phase diagram. Two state points with density and temperature $(\rho_1,T_1)$ and $(\rho_2,T_2)$ are termed {\it isomorphic} \cite{IV} if all pairs of physically relevant microconfigurations of the state points that trivially scale into one another (i.e., $\rho_1^{1/3}{\bf r}_i^{(1)}=\rho_2^{1/3}{\bf r}_i^{(2)}$ for all particles $i$), have proportional configurational Boltzmann factors:

\begin{equation}\label{isodef}
e^{-U({\bf r}_1^{(1)},...,{\bf r}_N^{(1)}) /k_BT_1}=C_{12}\,e^{-U({\bf r}_1^{(2)},...,{\bf r}_N^{(2)})/k_BT_2}\,.
\end{equation}
Only inverse-power law liquids \cite{ipl} have exact isomorphs (here $C_{12}=1$), but as shown in Appendix A of Ref. \onlinecite{IV} a system is strongly correlating if and only if it has isomorphs to a good approximation.

The invariance of the canonical probabilities of scaled microconfigurations along an isomorph has several implications, for instance \cite{scl_2,scl,IV}: 1) The excess entropy and the isochoric specific heat are isomorph invariants, 2) the reduced-unit dynamics is isomorph invariant for both Newtonian and stochastic dynamics, 3) all reduced-unit static correlation functions are isomorph invariant, 4) a jump between isomorphic state points takes the system instantaneously to equilibrium. Using reduced units means measuring length in terms the unit $\rho^{-1/3}$ where $\rho\equiv N/V$ is the particle density, and time in units of $\rho^{-1/3}\sqrt{m/k_BT}$ where $m$ is the average particle mass. Since isomorphs are generally approximate, isomorph properties are likewise rarely rigorously obeyed.

All thermodynamic quantities considered below are excess quantities, i.e., in excess of those of an ideal gas at the same density and temperature. Thus $S$ is the {\it excess} entropy ($S<0$), $C_V$ is the {\it excess} isochoric specific heat, $p$ is the {\it excess} pressure (i.e., $p=W/V$), etc.  

Briefly, the reason that $S$ and $C_V$ are isomorph invariants is the following \cite{IV}. The entropy is determined by the canonical probabilities, which are identical for scaled microconfigurations of two isomorphic state points. From Einstein's formula $C_V=\langle( \Delta U)^2\rangle/k_BT^2$ the isomorph invariance of $C_V$ follows easily by taking the logarithm of Eq. (\ref{isodef}) and making use of the isomorph invariance of scaled microconfiguration probabilities.

Since $S$ and $C_V$ are invariant along the same curves in the phase diagram, $C_V$ is a function of $S$: $C_V=\phi(S)$. Thus  $T(\partial S /\partial T)_V=\phi(S)$ or at constant volume: $\mathrm{d}S/\phi(S)=\mathrm{d}T/T$. Integrating this leads to an expression of the form $\psi(S)=\ln (T)+k(V)$, which implies $T=\exp[\psi(S)]\exp[-k(V)]$. The generic version of this involves only intensive quantities ($s\equiv S/N$):

\begin{equation}\label{fund1}
T
\,=\,f(s)h(\rho)\,.
\end{equation}
For inverse power law interactions ($\propto r^{-n}$) the entropy is well-known to be a function of $\rho^\gamma/T$ where $\gamma=n/3$: $S=K(\rho^\gamma/T)$. Applying the inverse of the function $K$, shows that these perfectly correlating systems obey Eq. (\ref{fund1}) with $h(\rho)=\rho^\gamma$.

The thermodynamic separation identity Eq. (\ref{fund1}) is the main result of this paper. We proceed to discuss some consequences and numerical tests.

{\it 1. Density scaling} 

\noindent Since entropy is an isomorph invariant, it follows from Eq. (\ref{fund1}) that the variable characterizing an isomorph may be chosen as $h(\rho)/T$. In particular, the reduced relaxation time $\tilde\tau$, which is also isomorph invariant, may be written for some function $G$

\begin{equation}\label{tau}
\tilde\tau=G\left(\frac{h(\rho)}{T}\right)\,.
\end{equation}
This is the form of  ``density scaling'' proposed by Alba-Simionesco {\it et al.} in 2004 from different arguments \cite{alb04}; at the same time Dreyfus {\it et al.}, as well as Casalini and Roland, favored the more specific form $\tilde\tau=G({\rho^\gamma}/{T})$ \cite{alb04}. Isochrones for many supercooled liquids and polymers follow to a good approximation the latter ``power-law density scaling'' relation \cite{densscal}. For large density changes, however, it was recently shown that the density-scaling exponent generally varies in both simulations and experiment \cite{lasse}; these cases conform to the more general Eq. (\ref{tau}).

{\it 2. An expression for the density-scaling exponent} 

\noindent The general, state-point dependent density-scaling exponent $\gamma$ is defined \cite{scl,IV} by

\begin{equation}\label{12}
\gamma
 \,\equiv\,\left(\frac{\partial \ln T}{\partial \ln\rho}\right)_S
\,=\,\left(\frac{\partial \ln T}{\partial \ln\rho}\right)_{\tilde \tau}
\,.
\end{equation}
The physical interpretation of Eq. (\ref{12}) is the following. If density is increased by 1\%, temperature should be increased by $\gamma$\% for the system to have the same entropy and reduced relaxation time. Equation (\ref{fund1}) implies $\mathrm{d}\ln T = \mathrm{d}\ln f(s) + \mathrm{d}\ln h(\rho)$; thus along an isomorph -- where $s$ and $\tilde\tau$ are both constant -- one has $\mathrm{d}\ln T = \mathrm{d}\ln h$. Via Eq. (\ref{12}) this implies 

\begin{equation}\label{gamma}
\gamma
 \,=\,\frac{\mathrm{d} \ln h}{\mathrm{d} \ln\rho}\,.
\end{equation}
In particular, $\gamma$ depends only on density: $\gamma =\gamma(\rho)$ \cite{IV}.

{\it 3. Configurational Gr{\"u}neisen equation of state} 

\noindent The Gr{\"u}neisen equation of state expresses that pressure equals a density-dependent number times energy plus a term that is a function of density only \cite{grun}. This equation of state is used routinely for describing, in particular, solids under high pressure. We proceed to show that strongly correlating systems obey the configurational version of the Gr{\"u}neisen equation of state, which as suggested by Casalini {\it et al.} \cite{cas06} has the density-scaling exponent as the proportionality constant \cite{IV,V}

\begin{equation}\label{gr}
W\,=\,\gamma(\rho)U+\Phi(\rho)\,.
\end{equation}
To prove this, note first that $\left(\partial U/\partial S\right)_\rho=T=f(S)h(\rho)$ by integration implies $U=F(S)h(\rho)+k(\rho)$ where $F'(S)=f(S)$ ($S$ is the extensive entropy). Since $W=\left(\partial U/\partial\ln\rho\right)_S$ (which follows from the standard identity $T\mathrm{d}S=\mathrm{d}U+p\mathrm{d}V$), we get $W=F(S)\mathrm{d}h/\mathrm{d}\ln\rho+\mathrm{d}k/\mathrm{d}\ln\rho$. Substituting into the latter expression $F(S)$ isolated from $U=F(S)h(\rho)+k(\rho)$ leads to Eq. (\ref{gr}), in which $\gamma(\rho)$ is given by Eq. (\ref{gamma}).

{\it 4. The isomorphs of atomic systems} 

\noindent We consider now predictions for systems of ``atomic'' particles interacting via pair potentials of the form \cite{rosen} (where $r$ is the distance between two particles)

\begin{equation}\label{v}
v(r)\,=\,\sum_n v_n r^{-n}\,.
\end{equation}
For simplicity of notation the case of identical particles is considered, but the arguments generalize trivially to multicomponent systems. Consider the thermal average $\langle r^{-n}\rangle$. Switching to reduced units defined by $\tilde r \equiv \rho^{1/3}r$, we have $\langle r^{-n}\rangle=\langle\tilde r^{-n}\rangle\rho^{n/3}$. Since structure is isomorph invariant in reduced units, $\langle\tilde r^{-n}\rangle$ is an isomorph invariant. Consequently, it is a function of any other isomorph invariant, for instance the entropy: $\langle\tilde r^{-n}\rangle=G_n(S)$. Noting that the average potential energy is a sum of Eq. (\ref{v}) over all particle pairs, we conclude that (where $H_n(S)\propto v_nG_n(S)$)

\begin{equation}\label{U}
U\,=\,\sum_n H_n(S) \rho^{n/3}\,.
\end{equation}
Taking the derivative of this equation with respect to temperature at constant volume leads to

\begin{equation}\label{cveq}
\left(\frac{\partial U}{\partial T}\right)_V
\,=\,\sum_n H'_n(S)\left(\frac{\partial S}{\partial T}\right)_V \rho^{n/3}\,.
\end{equation}
The left hand side is $T\left({\partial S}/{\partial T}\right)_V$, so Eq. (\ref{cveq}) implies

\begin{equation}\label{Tlig}
T\,=\,\sum_n H'_n(S)\rho^{n/3}\,.
\end{equation}
This is consistent with the thermodynamic separation identity Eq. (\ref{fund1}) only if all the functions $H'_n(S)$ are proportional to some function, i.e., if one can write $H'_n(S)=C_n \phi(S)$. We identify $\phi(S)$ as the function $f(s)$ of Eq. (\ref{fund1}), which means that

\begin{equation}\label{glig}
h(\rho)\,=\,\sum_n C_n\rho^{n/3}\,.
\end{equation}
Thus for strongly correlating atomic liquids, the thermodynamic function $h(\rho)$ has an analytical structure, which is inherited from $v(r)$ in the sense that the only non-zero terms of $h(\rho)$ are those corresponding to the non-zero terms of $v(r)$. Note that not all systems with potentials of the form Eq. (\ref{v}) are strongly correlating and that the derivation applies only if this is the case.

\begin{figure}[H]
  \centering
  \includegraphics[width=80mm]{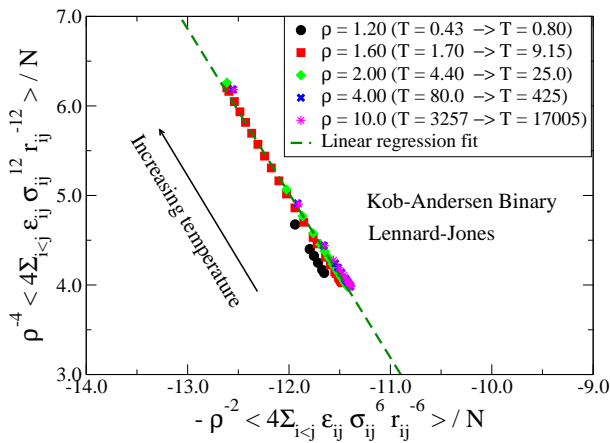}
  \caption{The thermal average of $r^{-12}$ versus that of $-r^{-6}$ in reduced units for a large range of state points of the Kob-Andersen binary Lennard-Jones liquid simulated with $1000$ particles ($\varepsilon_{AA}=\sigma_{AA}=1$). These quantities correspond to $H_{12}(S)$ and $H_{6}(S)$ in Eq. (\ref{U}). The theory predicts that $H'_{12}(S)\propto H'_{6}(S)$, implying that all data points should fall onto a common line according to $H_{12}(S)=\alpha H_{6}(S)+\beta$.}
  \label{fig1}
\end{figure}

As an illustration we present results from simulations of the Kob-Andersen binary Lennard-Jones (KABLJ) liquid \cite{ka}, which is strongly correlating at its condensed-phase state points \cite{scl_2,scl,IV}. The application of the above to LJ systems predicts that $H'_{12}(S)\propto H'_{6}(S)$, where $H_{12}(S)$ is the reduced coordinate average of the $r^{-12}$ term of $U$, etc. Integrating this leads to  $H_{12}(S)=\alpha H_{6}(S)+\beta$, implying that if the repulsive term in $U$ is plotted against the attractive term in reduced units, all points should fall onto a common line. Figure \ref{fig1} presents  data where density was changed by a factor of eight and temperature a factor of 40,000. The data collapse is good but not exact, which reminds us that the relations derived are approximate.

The theory implies a simple mathematical description of the isomorphs in the $(\rho,T)$ phase diagram. From the fact that the potential energy contains only $r^{-12}$ and $r^{-6}$ terms, it follows that $h(\rho)=A \rho^4 - B\rho^2$. Consequently, LJ isomorphs are given by

\begin{equation}\label{ljisom}
\frac{A\rho^4 -B\rho^2}{T}\,=\,{\rm Const.}
\end{equation}
The invariance of the Boltzmann statistical weights of scaled microconfigurations implies that an isomorph cannot cross the liquid-solid coexistence curve. In particular, the coexistence curve is itself predicted to be an isomorph \cite{IV}, which was recently confirmed by simulations of generalized LJ liquids \cite{V,sadus}. Consequently the coexistence line is given by Eq. (\ref{ljisom}). This validates a recent conjecture of Khrapak and Morfill \cite{khr11}.

{\it 5. Predictions for the repulsive Lennear-Jones fluid} 

\noindent As a final illustration we consider the ``repulsive'' single-component LJ fluid defined by the pair potential $(r^{-12}+r^{-6})/2$, a system with $WU$ correlation coefficient above 99.9\% in its entire phase diagram. At low densities ($\rho\ll 1$) the repulsive LJ fluid behaves as an $r^{-6}$ fluid, whereas it for $\rho\gg 1$ is effectively an $r^{-12}$ fluid. Thus the density-scaling exponent $\gamma(\rho)$ varies from $2$ to $4$ as density increases, a much larger variation than that of previously studied strongly correlating systems. 

\begin{figure}[H]
  \centering
  \includegraphics[width=80mm]{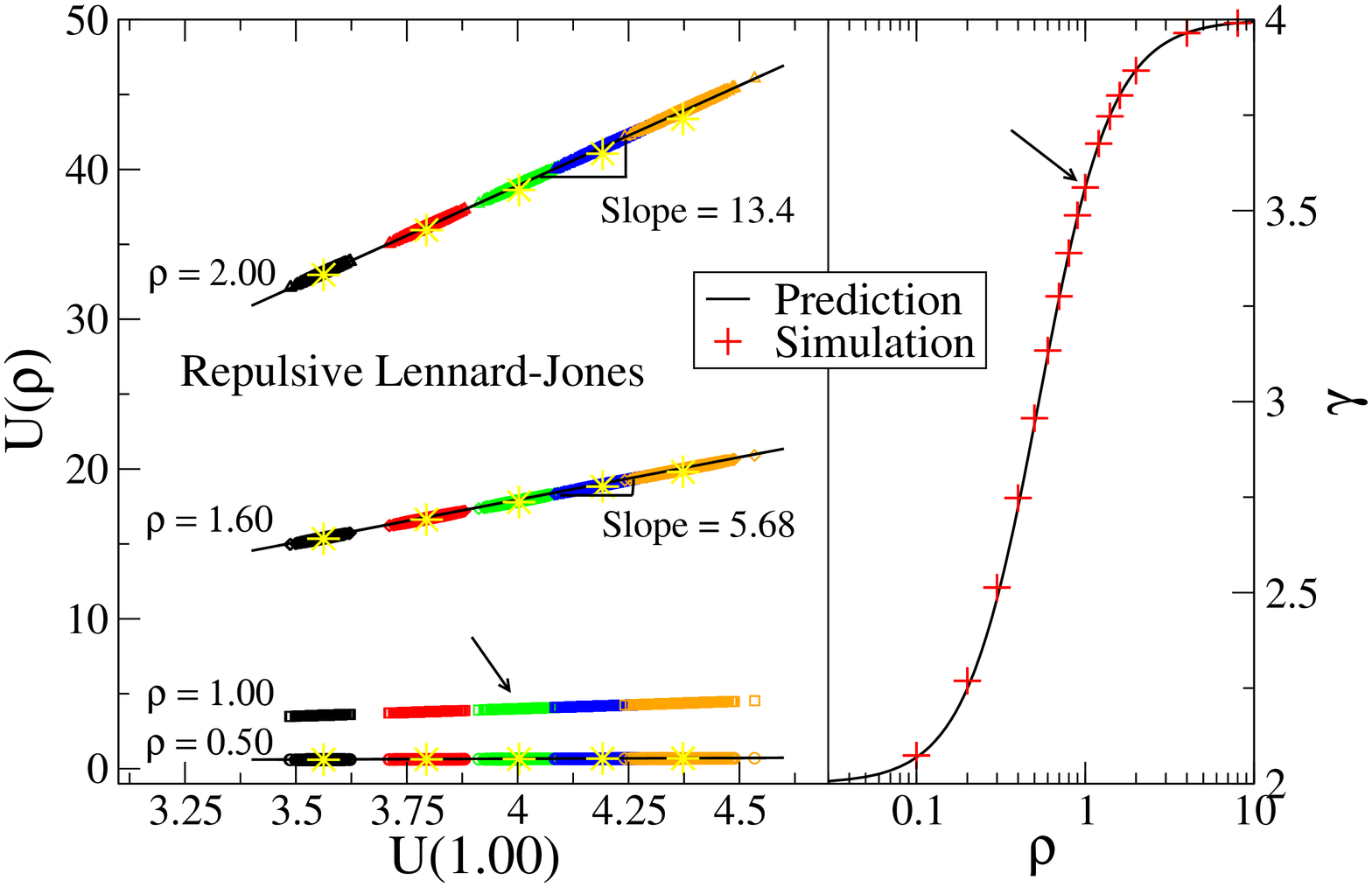}
  \caption{``Multiple direct isomorph check'' applied to simulations of $N=1000$ particles of the repulsive LJ fluid defined by the pair potential $(r^{-12}+r^{-6})/2$. The left panel shows the potential energies of pairs of microconfigurations, where the potential energy of a given microconfiguration at density $1.0$ is denoted $U(1.00)$ and that of the same microconfiguration scaled to density $\rho$ is denoted $U(\rho)$ ($\rho=0.5; 1.6; 2.0$). This was done for $T=0.6; 0.8; 1.0; 1.2; 1.4$. The black lines are the predictions (see the text) with slopes determined via Eq. (\ref{g6_12}) from the fluctuations calculated at the state point $(\rho,T)=(1,1)$ marked by an arrow. The right panel shows the density-scaling exponent for each state point along an isomorph predicted from Eqs. (\ref{gamma}) and (\ref{g6_12}) (full curve) and the exponent  calculated via the fluctuation formula $\gamma=\langle\Delta W\Delta U\rangle/\langle(\Delta U)^2\rangle$ \cite{IV} (red crosses). The arrow marks the state point $(\rho,T)=(1,1)$.}
  \label{fig2}
\end{figure}

Since $h(\rho)$ is only defined within an overall multiplicative constant, one can write for the repulsive LJ fluid $h(\rho)=\alpha \rho^4+(1-\alpha)\rho^2$. This leads via Eq. (\ref{gamma}) to $\gamma_0=2+2\alpha$, implying that  

\begin{equation}\label{g6_12}
h(\rho)\,=\,(\gamma_0/2-1) \rho^4+(2-\gamma_0/2)\rho^2\,.
\end{equation}
Our simulations identified from the expression $\gamma_0=\langle\Delta W\Delta U\rangle/\langle(\Delta U)^2\rangle$ \cite{IV} the exponent $\gamma_0=3.56$ at the state point $(\rho,T)=(1,1)$. Equation (\ref{g6_12}) with $\gamma_0=3.56$ was tested in two different ways. First, we compared at each state point along an isomorph the exponent $\gamma(\rho)$ predicted from Eqs. (\ref{gamma}) and (\ref{g6_12}) with that calculated from the fluctuations via $\gamma=\langle\Delta W\Delta U\rangle/\langle(\Delta U)^2\rangle$ (right panel of Fig. 2). The left panel presents a second test of Eq. (\ref{g6_12}) by showing results from simulating five temperatures at $\rho=1$, plotting for each temperature instantaneous values of the potential energy versus the potential energy of the same microconfigurations scaled to three other densities ($\rho=0.5, 1.6, 2.0$). The theory behind the observed straight lines is the following. Consider two isomorphic state points $(\rho_0,T_0)$ and $(\rho,T)$ and suppose each temperature is changed a little, keeping both densities constant. If the two new state points are also isomorphic, the entropy change is the same for both: ${dU_0}/{T_0}={dU}/{T}$. This implies $dU/dU_0=T/T_0$, i.e., $\left({\partial U}/{\partial U_0}\right)_{\rho_0,\rho}=T/T_0$. Since $h(\rho)/T$ is constant along an isomorph, this implies $\left({\partial U}/{\partial U_0}\right)_{\rho_0,\rho}=h(\rho)/h(\rho_0)$. Integrating this at constant $\rho_0$ and $\rho$ leads to $U=[{h(\rho)}/{h(\rho_0)}]U_0+\phi(\rho_0,\rho)$. In our case $\rho_0=1$ and $h(\rho_0)=1$. Thus plotting $U$ versus $U_0$ is predicted to result in straight lines with slope $h(\rho)$ (yellow asterices in the left panel of Fig. \ref{fig2}). The scaled state points are isomorphic to the original $\rho=1$ state points, with temperatures given by $T=T_0h(\rho)$. Via the ``direct isomorph check'' \cite{IV} this implies that the scaled microconfigurations form elongated ovals with slope $h(\rho)$.

In summary, we have shown that for strongly correlating liquids or solids, temperature separates into a function of entropy times a function of density. For these systems the energy scale is consequently determined by density alone. It is an open question whether, conversely, the thermodynamic separation identity Eq. (\ref{fund1}) implies that the system in question is strongly correlating.
\vspace{1cm}
%\acknowledgments 

The centre for viscous liquid dynamics ``Glass and Time'' is sponsored by the Danish National Research Foundation (DNRF).


\begin{thebibliography}{99}

\bibitem{scl_2}
N. P. Bailey {\it et al.}, J. Chem. Phys. {\bf 129}, 184507 (2008);
N. P. Bailey {\it et al.}, J. Chem. Phys. {\bf 129}, 184508 (2008);
T. B. Schr{\o}der {\it et al.}, J. Chem. Phys. {\bf 131}, 234503 (2009).

\bibitem{scl}
U. R. Pedersen {\it et al.}, Phys. Rev. Lett. {\bf 100}, 015701 (2008);
N. Gnan {\it et al.}, Phys. Rev. Lett. {\bf 104}, 125902 (2010);
U. R. Pedersen {\it et al.}, Phys. Rev. Lett. {\bf 105}, 157801 (2010).

\bibitem{IV}
N. Gnan {\it et al.}, J. Chem. Phys. {\bf 131}, 234504 (2009).

\bibitem{V}
T. B. Schr{\o}der {\it et al.}, J. Chem. Phys.  {\bf 134}, 164505 (2011).

\bibitem{rome}
U. R. Pedersen {\it et al.}, J. Non-Cryst. Solids {\bf 357}, 320 (2011).

\bibitem{ditte}
D. Gundermann {\it et al.}, Nature Physics {\bf 7}, 816 (2011).

\bibitem{simple}
T. S. Ingebrigtsen, T. B. Schr{\o}der, and Jeppe C. Dyre, arXiv:1111.3557 (2011).

\bibitem{liq}
O. Hirschfelder, C. F. Curtiss, and R. B. Bird, {\it Molecular theory of gases and liquids} (Wiley, New York, 1954);
J. P. Boon and S. Yip, {\it  Molecular hydrodynamics} (McGraw-Hill, New York, 1980);
J. S. Rowlinson and B. Widom, {\it Molecular theory of capillarity} (Clarendon, Oxford, 1982); 
M.P. Allen and D.J. Tildesley, {\it Computer simulation of liquids} (Oxford Science Publications, Oxford, 1987);
D. Chandler, {\it Introduction to modern statistical mechanics} (Oxford University Press, New York, 1987); 
P. G. Debenedetti, {\it Metastable liquids: Concepts and principles} (Princeton University Press, Princeton, NJ, 1996);
N. H. March and M. P. Tosi, {\it Introduction to liquid state physics} (World Scientific Publishing, Singapore, 2002);
J.-L. Barrat and J.-P. Hansen, {\it Basic concepts for simple and complex liquids} (Cambridge University Press, Cambridge, England, 2003);
J.-P. Hansen and J. R. McDonald, {\it Theory of simple liquids}, 3rd ed. (Academic, New York, 2005).

\bibitem{ipl}
O. Klein, Medd. Vetenskapsakad. Nobelinst. {\bf 5}, 1 (1919);
T. H. Berlin and E. W. Montroll, J. Chem. Phys. {\bf 20}, 75 (1952);
W. G. Hoover {\it et al.}, J. Chem. Phys. {\bf 52}, 4931 (1970);
W. G. Hoover, S. G. Gray, and K. W. Johnson, J. Chem. Phys. {\bf 55}, 1128 (1971);
Y. Hiwatari {\it et al.}, Prog. Theor. Phys. {\bf 52}, 1105 (1974);
D. M. Heyes and A. C. Branka, J. Chem. Phys. {\bf 122}, 234504 (2005);
A. C. Branka and D. M. Heyes, Phys. Rev. E {\bf 74}, 031202 (2006).

\bibitem{alb04}
C. Alba-Simionesco, D. Kivelson, and G. Tarjus, J. Chem. Phys. {\bf 116}, 5033 (2002);
C. Dreyfus {\it et al.}, Phys. Rev. E {\bf 68}, 011204 (2003);
C. Alba-Simionesco {\it et al.}, Europhys. Lett. {\bf 68}, 58 (2004);
R. Casalini and C. M. Roland, Phys. Rev. E {\bf 69}, 062501 (2004).


\bibitem{densscal}
C. M. Roland {\it et al.}, Rep. Prog. Phys. {\bf 68}, 1405 (2005);
G. Floudas, M. Paluch, A. Grzybowski, and K. L. Ngai, {\it Molecular Dynamics of Glass-Forming Systems: Effects of Pressure} (Advances in Dielectrics, Springer, 2010);
D. Fragiadakis and C. M. Roland, J. Chem. Phys. {\bf  134},  044504 (2011).

\bibitem{lasse} L. B{\o}hling {\it et al.}, arXiv:1112.1602 (2011).

\bibitem{grun}
M. Born and K. Huang, {\it Dynamical Theory of Crystal Lattices} (Oxford University Press, Oxford U.K., 1954);
M. Ross and D. A. Young, Annu. Rev. Phys. Chem. {\bf 44}, 61 (1993);
L. Burakovsky and D. L. Preston, J. Phys. Chem. Solids {\bf 65}, 1581 (2004).

\bibitem{cas06} R. Casalini, U. Mohanty, and C. M. Roland, J. Chem. Phys. {\bf 125}, 014505 (2006).

\bibitem{rosen}
Y. Rosenfeld, Phys. Rev. A {\bf 26}, 3633 (1982).

\bibitem{ka}
W. Kob and H. C. Andersen, Phys. Rev. Lett. {\bf 73}, 1376 (1994).

\bibitem{sadus}
A. Ahmed and R. J. Sadus, J. Chem. Phys. {\bf 131}, 174504 (2009).

\bibitem{khr11}
S. A. Khrapak and G. E. Morfill, J. Chem. Phys. {\bf 134}, 094108 (2011).

\end{thebibliography}
\end{document}